\documentclass[column,preprintnumbers,nofootinbib,prl]{revtex4}
\usepackage{graphicx}

\usepackage[hypertex]{hyperref}
\newcommand{\beq}{\begin{equation}}
\newcommand{\eeq}{\end{equation}}
\newcommand{\be}{B_\oplus}

\def\be{\begin{equation}}
\def\ee{\end{equation}}
\def\baray{\begin{eqnarray}}
\def\earay{\end{eqnarray}}
\def\ba{\begin{eqnarray}}
\def\ea{\end{eqnarray}}

\begin{document}

\title{\textbf{Zeeman Effect In The Framework of \\Moyal Noncommutativity and String Theory}}

\author{\textbf{M.B. Sedra, A. El Boukili}}
\affiliation{International Centre for Theoretical Physics, Trieste,
Italy,\\ Groupement National de Physique de Hautes Energies, GNPHE,
Rabat, Morocco,\\ Laboratoire de Physique de La Mati\`ere et
Rayonnement (LPMR), Univ. Ibn Tofail, Facult\'e des Sciences,
K\'{e}nitra, Morocco}

\author{\textbf{E.H. Saidi}}
\affiliation{International Centre for Theoretical Physics, Trieste,
Italy,\\ Groupement National de Physique de Hautes Energies, GNPHE,
Rabat, Morocco,\\ Laboratoire de Physique des Hautes Energies
(LPHE), Univ. $M^{d}$. V, Agdal, Facult\'e des Sciences, Rabat,
Morocco}

\begin{abstract}
\medskip Stimulated by the importance of noncommutative geometry in recent
developments in string theory, D-branes and integrable systems;
one intends in this work to present a new insight towards adapting
the famous idea of Zeeman effect to noncommutativity \`a la Moyal
and develop an analysis leading to
connect our results to the Bigatti-Suskind (BS) formulation.\\\\
\end{abstract}

%\keywords {Zeeman effect, Integrable systems, String and Branes}
\maketitle
\newpage
\section{Introduction}

Noncommutative geometry (NCG) or simply noncommutativity of
coordinates, is a very old idea \cite{snyder}. The use of NCG in
string theory was initiated for the first time by Witten
\cite{witten} and recently, through the original work carried out
by Connes, Douglas and Schwartz \cite{1, CDS}, this idea took a
new dash in mathematics and physics. Several important
contributions on NCG followed these seminal works \cite{5, 6, 7},
we refer the readers to the textbooks \cite{books} for general
references on NCG and string theory.\newline

We are focusing to contribute to this very relevant research
subject by using the Moyal momentum noncommutativity \cite{BSed,
BSed2} and its various applications in $2d$ integrability and
conformal symmetry. We suggest then an interpretation of the
noncommutative ($sl_{2}$ KdV and $sl_{3}$ Boussinesq)-Burger's
mappings \cite{DES, Sed} by implementing the famous idea of Zeeman
effect.\newline

This interpretation starts to emit signs of rigor and reliability
by means of the BS construction \cite{BSus} which we are engaging
in our approach. The original idea to incorporate the Zeeman
effect in the formalism of Moyal noncommutativity in connection
with certain aspects of $2d$ integrable models and string-brane
theories requires, undoubtedly, a particular interest. More
research's works in this direction will be focused in future
occasions.

\section{Basic definitions}

Let's first start by specifying the nature of objects used in this
work. The functions often involved in the two dimensional
phase-space are arbitrary
functions which we generally indicate by $f(x,p)$ with coordinates $x$ and $%
p $. With respect to this phase space, we have to define the
following objects:

{\bf 1}. The constants $f_0$ defined such that
\begin{equation}
\partial_x f_0=0= \partial_p {f_0}.
\end{equation}

{\bf 2}. The functions ${u_i}(x,t)$ depending on an infinite set
of variables $t_1 =x, t_2,t_3,...,$ with
\begin{equation}
\partial_{p} u_{i} (x,t)=0.
\end{equation}
The index $i$, stands for the conformal weight of the field $u_i
(x,t)$. These functions can be considered in the complex language
framework as being the analytic (conformal) fields of conformal
spin $i=1,2,...$.

{\bf 3}. Other objects that we will use are given by
\begin{equation}
u_{i}(x,t)\star p^{j}
\end{equation}
These are objects of conformal weight $(i+j)$ living on the
non-commutative space parametrized by $\theta $. Throughout this
work, we will use the
following convention notations: $[u_{i}]=i$, $[\theta ]=0$, and $%
[p]=[\partial _{x}]=-[x]=1$, where the symbol $[\hspace{0cm}$ $]$
stands for the conformal dimension of the enclosed object.

{\bf 4}. The star product law defining the multiplication of
objects in the non-commutative space is shown to satisfy the
following expression
\begin{equation}
f(x,p)\star g(x,p)= \sum_{s=0}^{\infty}
\sum_{i=0}^{s}{\frac{\theta ^s}{s!}} (-)^{i} c _{s}^{i}
(\partial_{x}^{i}\partial_{p}^{s-i}f)(\partial_{x}^{s-i}\partial_{p}^{i}g)
\end{equation}
with $c _{s}^{i}=\frac {s!}{i!(s-i)!} $.

{\bf 5}. The Moyal bracket is defined as
\begin{equation}
\{f(x,p),g(x,p)\}_{\theta }=\frac{f\star g-g\star f}{2\theta }.
\end{equation}

{\bf 6}. In order to distinguish the classical objects from the $\theta $%
-deformed ones, we consider the following convention
notations:\newline a) $\widehat{\Sigma }_{m}^{(r,s)}$: is the
space of momentum Lax differential operators of conformal spin $m$
and degrees $(r,s)$ with $r\leq s$. Typical operators of this
space are given by
\begin{equation}
\sum_{i=r}^{s}u_{m-i}\star p^{i}.
\end{equation}%
\newline
b) $\widehat{\Sigma }_{m}^{(0,0)}$: Is the space of coefficient functions $%
u_{m}$ of conformal spin $m$; $m\in Z$, which may depend on the parameter $%
\theta $. It coincides in the standard limit, $\theta =0$, with
the ring of
analytic fields involved in the construction of conformal symmetry and $W$%
-extensions.\newline c) ${\widehat{\Sigma }}_{m}^{(k,k)}$: Is the
space of momentum operators of type
\begin{equation}
u_{m-k}\star p^{k}.
\end{equation}%
{\bf 7}. {\bf The\ } $\theta $-{\bf Leibnitz rules}\cite{BSed}
\begin{equation}
p^{n}\star f(x,p)=\sum_{s=0}^{n}\theta
^{s}c_{n}^{s}f^{(s)}(x,p)p^{n-s},
\end{equation}%
and
\begin{equation}
p^{-n}\star f(x,p)=\sum_{s=0}^{\infty }(-)^{s}\theta
^{s}c_{n+s-1}^{s}f^{(s)}(x,p)p^{-n-s},
\end{equation}%
where $f^{(s)}=\partial _{x}^{s}f$ is the prime derivative. Few
examples are given by

\begin{equation}
\begin{array}{lcl}
1\star f(x,p) & = & f, \\
p\star f(x,p) & = & fp+\theta f^{\prime }, \\
p^{2}\star f(x,p) & = & fp^{2}+2\theta f^{\prime }p+{\theta
}^{2}f^{\prime
\prime }, \\
p^{3}\star f(x,p) & = & fp^{3}+3\theta f^{\prime 2}+3\theta
^{2}f^{\prime
\prime }p+{\theta }^{3}f^{\prime \prime \prime },%
\end{array}%
\end{equation}

\bigskip and
\begin{equation}
\begin{array}{lcl}
p^{-1}\star f(x,p) & = & fp^{-1}-\theta f^{\prime -2}+{\theta
}^{2}f^{\prime
\prime -3}-{\theta }^{3}f^{\prime \prime \prime -4}+..., \\
p^{-2}\star f(x,p) & = & fp^{-2}-2\theta f^{\prime -3}+3{\theta }%
^{2}f^{\prime \prime -4}-4{\theta }^{3}f^{\prime \prime \prime -5}+..., \\
p^{-3}\star f(x,p) & = & fp^{-3}-3\theta f^{\prime -4}+6{\theta }%
^{2}f^{\prime \prime -5}-10{\theta }^{3}f^{\prime \prime \prime -6}+....%
\end{array}%
\end{equation}

\bigskip {\bf 8. The Ring of analytic functions:}

A convenient description consists in using the complex language
notation in which we define the two dimensional Euclidean space
parametrized by $z=t+ix$ and ${\bar{z}}=t-ix$. In this notation,
the currents of conformal weights $k$ are simply written as
$u_{k}(x,t)\equiv u_{k}(z)$. \newline It's then the time to
introduce the space of analytic functions of arbitrary conformal
spin. This is the space of completely reducible infinite
dimensional $so(2)$ Lorentz representations that can be written as
\begin{equation}
\Sigma ^{(0,0)}=\oplus _{k\in Z}\Sigma _{k}^{(0,0)},
\end{equation}%
where the $\Sigma _{k}^{(0,0)}$'s are one dimensional $so(2)$ spin
$k$ irreducible modules. The upper indices $(0,0)$ carried by the
space $\Sigma ^{(0,0)})$ are special values of the general indices
$(p,q)$ describing the lowest and highest degrees of Lax operators
type $\sum_{i=p}^{q}u_{m-i}\star p^{i}.$The generators belonging
to the space $\Sigma ^{(0,0)})$ are given by the spin $k$ analytic
fields. They may be viewed as analytic maps $u_{k}$ which
associate to each point $z$ on the unit circle the fields
$u_{k}(z)$. For $k\geq 2$, these fields can be thought of as the
higher spin currents involved in the construction of $w_{\theta
}$-algebras. The particular example is given by the spin-2 current
$u_{_{2}}(z)$ intimately associated to the Virasoro algebra
$T(z)$.

\bigskip

{\bf 9. The classical limit}

Since we are interested in the $\theta $-deformation case, we have
to add that the spaces $\Sigma _{k}^{(0,0)}$ are $\theta
$-dependent; the corresponding $w_{\theta }$-algebra is shown to
exhibit new properties related to the $\theta $-parameter and
reduces to the standard $w$-algebra once some special limits are
performed. As an example, consider for instance the $w_{\theta
}^{3}$-algebra generalizing the Zamolodchikov algebra. The
conserved currents of this extended algebra are shown to take the
following form
\begin{equation}
\begin{array}{lcl}
w_{2} & = & u_{2}, \\
w_{3} & = & u_{3}-\theta u_{2}^{^{\prime }},%
\end{array}%
\end{equation}%
which coincides with the classical case once the limit $\theta
=\frac{1}{2}$ is performed. It is the convenient limit than we
must consider in field theory and integrable systems to assure
compatibility with the extended conformal symmetry (Zamolodchikov
algebra) since the standard limit $\ \theta =0$ doesn't respect
this objective

\section{Zeeman Effect and the Burger-$sl_{n}$ KdV Mapping}

We present in this section an original approach to interpret some
results previously established and that concern the mapping
between the NC Burger's integrable system and the NC deformation
of the $sl_{n}$-KdV integrable hierarchies in the Moyal momentum
space \cite{BSed, BSed2, DES, Sed}. This approach consists in
adapting, artfully, the famous idea of Zeeman effect to the Moyal
algebra of NC Lax operators to provide an alternative issue and
interpretation of some properties that are encountered in the
study of NC integrable hierarchies.\newline

Let's underline in this context that the presence of the
deformation parameter $\theta $ in the Moyal algebra is important
in the sense that it
can leads to identify it with the inverse of the magnetic field $B$, as it%
%TCIMACRO{\U{b4}}%
%BeginExpansion
\'{}%
%EndExpansion
s well known, such that
\begin{equation}
\theta \sim B^{-1}
\end{equation}
Before considering such an application, it's useful to recall some
essential basic notions of the Zeeman effect as well as the
importance of the magnetic field in this context

\subsection{Zeeman Effect: Basic ideas}

\subsubsection{Definitions:}

The following definitions are equivalents:\newline {\bf 1.
}Knowing that an atom can be characterized by a unique set of
discrete energy states, when excited, the atom makes transitions
between these quantized energy states and emits light. The emitted
light is shown to form a discrete spectrum, reflecting the
quantized nature of the energy levels. In the presence of a
magnetic field, these energy levels can shift, this is the Zeeman
effect.\newline {\bf 2.} Analogously to the Stark effect
characterized by the splitting of a spectral line into several
components in the presence of an electric field, the Zeeman effect
is defined as been the splitting of a spectral line into several
components in the presence of a magnetic field.

\subsubsection{Origin of the Zeeman Effect:}

The origin of the Zeeman effect can be simply presented as
follows:\newline Let's consider an atomic energy state such that
an electron orbits around the nucleus of the atom. This electron
has a magnetic dipole moment associated with its angular momentum.
In the presence of a magnetic field, the electron acquires an
additional energy and consequently the original energy level is
shifted. The energy shifted may be positive, zero, or even
negative, depending on the angle between the electron magnetic
dipole moment and the field.\newline

\subsection{Zeeman Effect And The Burger - $Sl_{n}$ KdV Mapping}

We have shown in previous works \cite{DES, Sed} how it's possible
to establish a correspondence between the $sl_{n}$-KdV NC
integrable hierarchies and the NC Burger's system. This issue of
mapping exhibits a particular interest because it allows us to
install the first steps toward a
possible unification of NC integrable models in the Moyal momentum framework.%
\newline

Presently, we are interested to study another aspect in relation
with the Zeeman effect. The crucial point resides in the NC
deformation that provides the possibility to join the parameter of
non commutativity $\theta $ with the inverse of the magnetic field
$B$. On the basis of this relation, and also on the use of the NC
version of the Miura transformation which rests on the idea of
mapping between Burger and $sl_{n}$-KdV systems, we will see
explicitly how a strong analogy emerge between the Zeeman effect
and what the mapping in question exhibits as consequences.\newline

We guess that the incorporation of the Zeeman effect in this
context is not a coincidence, we think that some important
physical properties are behind. The starting steps in planting
this idea comes from several observations of the behavior of
different expressions and also from the primordial role of the NC
Burger's system and of the $\theta $ parameter whose weight
increases proportionally with the order of the $sl_{n}$-KdV
hierarchy.\newline

We are going to illustrate these ideas for two particular examples
namely the NC KdV and Boussinesq's systems.

\subsubsection{\protect\bigskip The $sl_{2}$ KdV- Burger's Case}

Let us take again the NC KdV-Burger's mapping discussed before, we
have the following equation
\begin{equation}
L_{KdV}(u_2)=p^{2}+u_2=(p+u_1)\star(p-u_1)
\end{equation}
or equivalently $u_2=-u_{1}^{2}-2\theta u^{\prime}_{1}$.

Whereas the NC KdV current $u_{2}$ depends explicitly on the parameter $%
\theta $ \footnote{%
Because of the fact that the NC KdV operator admits the NC Burger
Lax operator as a square root, see eq (2)}, the Burger's current
does not have this property since the associated Lax operator
$L_{Burger}(u_{1})$ does not admit a Lax operator, whose conformal
weight is integer, as a root. This is also due to the fact that
the quantities of non integer conformal weights are not authorized
in this framework.\newline

Thus from now on, the NC Burger's current is regarded as being a
fundamental current in term of which all the other currents of the
$sl_n$-KdV hierarchy are expressed.\newline

{\bf Proposition 1}:\newline

Given the NC Miura Transformation, binding the NC Burger and KdV
systems as follows:
\begin{equation}
L_{KdV}(u_{2})=L_{Burg}(u_{1})\star L_{Burg}(-u_{1}),
\end{equation}

We can represent this mapping graphically as given by \emph{fig.1}:
\begin{figure}[tbp]
\includegraphics[width=0.86\linewidth]{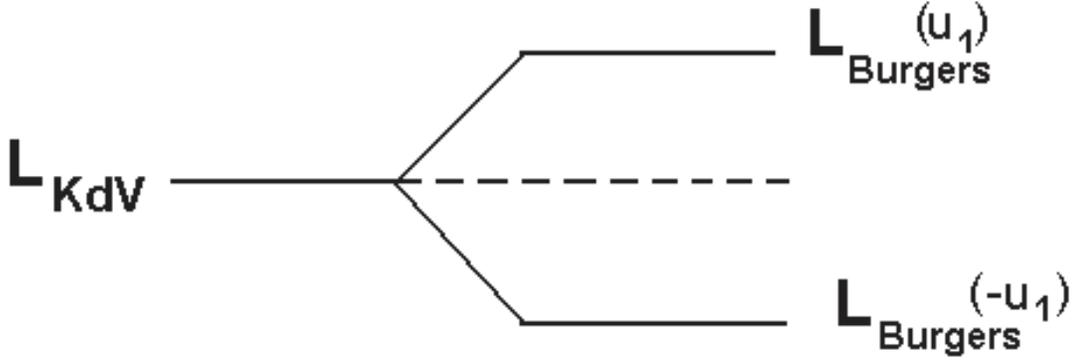}
\caption{\textbf{The $sl_2$-KdV-Burger mapping}} \label{fig1}
\end{figure}\\

{\bf Convention notations}:\newline

For this purpose, we adopt the following diagrammatic
representation:\newline {\bf 1} We represent symbolically the NC KdV
Lax operator by a line indexed by the NC KdV current $u_{2}$. The
splitting of $L_{KdV}$, with respect to the Moyal star product, into
a pair of NC Burger's operators is schematized by two parallel lines
which leave, starting from a vertex, the initial KdV line. The two
parallel lines are considered to be associated to the pair of NC
Burger's operators $(L_{Burg}, L_{Burg} )$ who appear on the right
hand side of the eq.(16).\newline

{\bf 2}. The position of the two emitted lines relative to the
Burger's operators depends on the sign of the spin 1 Burger's
currents. The upper line is associated with $L_{Burg}(+u_1)$ while
the lower one, associated to
negative sign of the NC Burger's current $-u_1$, is for $L_{Burg}(-u_1)$.%
\newline
{\bf 3}. The initial NC KdV line is thus the result of the star
product of the parallel Burger's lines.\newline

{\bf 4}. We then specify two zones; the left one, characterized by
the NC KdV initial line where the two Burger's levels are in
coincidence (degenerated twice). The other zone on the right is
given by two separated lines describing a broken degeneracy. The
passage from the single KdV level to both Burger's lines, via the
Miura transformation or the $sl_2$ KdV-Burger's mapping, is
identical to a lifting of the degeneracy which means also the
passage from a configuration with star product to a configuration
without star product:
\begin{equation}
L_{KdV}(u_{2})\hookrightarrow (L_{Burg}(u_{1}), L_{Burg}(-u_{1}) )
\end{equation}%
\newline
or equivalently
\begin{equation}
(p+u_1)\star(p-u_1)\hookrightarrow \left((p+u_1), (p-u_1)\right)
\end{equation}
{\bf 5}. In other words, the transition from a single KdV level
degenerated twice to a pair of two Burger's levels without
degeneracy is equivalent to the passage from a phase with a
$\theta \equiv B^{-1}$ predominance to a
phase with magnetic $B\equiv \theta ^{-1}$ predominance \footnote{%
At the level of the single NC KdV line, where the degeneracy is of
order 2,
the $\theta$ parameter acquires a power conversely to the magnetic field $%
\theta^{-1}$ which becomes relevant with emission of the pair of
NC Burger's levels}. At this point, we have to underline the
striking analogy with the Zeeman effect since it is the presence
of a magnetic field who breaks the degeneracy of the initial KdV
level.\newline

We showed through this first example that the KdV-Burger's mapping
is accompanied by a rupture of degeneracy that exhibits the
initial NC KdV level. This lifting of degeneracy is due to the
emergence of the magnetic
field $B$ corresponding to a weakness of the NC deformation parameter $%
\theta $ during the transition from the single KdV level to the
pair of Burger's levels. This behavior is identical to the Zeeman
effect.

\subsubsection{The $sl_{3}$ Boussinesq-Burger's Case}

In a similar way, the $sl_{3}$\ Boussinesq-Burger's mapping deals
with the following equation
\begin{equation}
p^{3}+u_{2}\star p+u_{3}=(p+u_{1})\star (p+v_{1})\star
(p-u_{1}-v_{1})
\end{equation}%
\newline
\newline
{\bf Proposition 2}:\newline

Given the NC Miura Transformation, binding the NC Burger and
$sl_3$ Boussinesq's system as follows
\begin{equation}
L_{Bouss}(u_{2}, u_3)=L_{Burg}(u_{1})\star L_{Burg}(v_{1})\star
L_{Burg}(-u_{1}-v_{1}) ,
\end{equation}

We can represent this mapping graphically as shown by \emph{fig.2} :
\begin{figure}[tbp]
\includegraphics[width=0.86\linewidth]{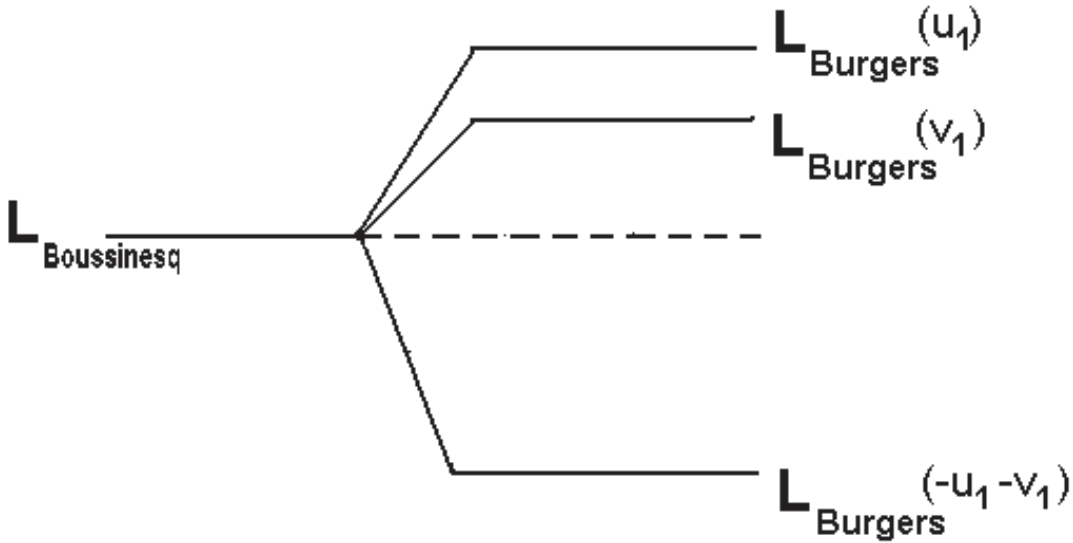}
\caption{\textbf{The $sl_3$-Boussinesq-Burger mapping}} \label{fig2}
\end{figure}\\

{\bf Convention notations}:\newline

In the same way as for the NC KdV- Burger's mapping, we will adopt
the following diagrammatic representations for the $sl_3$
Boussinesq-Burger's splitting:\newline

{\bf 1.} We represent symbolically the NC $sl_3$ Boussinesq Lax
operator by
a line indexed by the pair of currents $(u_{2}, u_{3})$. The splitting of $%
L_{Bouss}$, with respect to the star product, into a triplet $%
(L_{Burg}(u_{1}), L_{Burg}(v_{1}), L_{Burg}(-u_{1}-v_{1}))$ of NC
Burger's operators is schematized by three parallel lines leaving,
from a vertex, the initial $sl_3$ Boussinesq line.\newline

{\bf 2.} The position of the three emitted Burger's lines depends
on the sign of spin 1 Burger's currents. The uppers lines are
chosen arbitrarily to be associated to $L_{Burg}(+u_1)$ and
$L_{Burg}(+v_1)$ while the lower line
is being associated to the negative sign of the NC Burger's current $%
-u_{1}-v_{1}$ namely to $L_{Burg}(-u_{1}-v_{1})$.\newline

{\bf 3.} As it's shown in eq(19), the initial NC $sl_{3}$ Boussinesq
line is the result of the star product of the three parallel
Burger's lines.\newline

{\bf 4.} The initial single line corresponding to the NC $sl_3$
Boussinesq's Lax operator is characterized by a degeneracy of
order 3. This is due to the
fact that the three NC Burger's levels on the right of the vertex {\em fig. 2%
} coincide at the level of the NC $sl_3$ Boussinesq line. When the
degeneracy of order three is broken, the three NC Burger's levels
are emitted. This emission procedure is given by
\begin{equation}
L_{Bouss }(u_{2}, u_{3})\hookrightarrow (L_{Burg}(u_{1}),
L_{Burg}(v_{1}), L_{Burg}(-u_{1}-v_{1}))
\end{equation}
or equivalently
\begin{equation}
(p+u_{1})\star (p+v_{1})\star (p-u_{1}-v_{1})\hookrightarrow
((p+u_{1}), (p+v_{1}), (p-u_{1}-v_{1}))
\end{equation}
{\bf 5}. The contact with Zeeman effect is done as
follows:\newline
\newline
The transition from the initial single NC $sl_{3}$ Boussinesq's
level, of degeneracy three, to a triplet of NC Burger's levels
without degeneracy is equivalent to the passage from a phase with
a $\theta \equiv B^{-1}$ predominance to a phase with magnetic
$B\equiv \theta ^{-1}$ predominance. At this point, we have to
underline the striking analogy with the Zeeman effect since it is
the presence of a magnetic field who breaks the degeneracy of the
initial $sl_3$ Boussinesq's level.\newline
\newline
We showed once again through this second example that the $sl_{3}$
Boussinesq-Burger's mapping is accompanied by a rupture of
degeneracy, of order three, that exhibits the initial NC
Boussinesq's level. This lifting of degeneracy is due to the
emergence of the magnetic field $B\sim \theta
^{-1}$ accompanied by an annihilation of the NC deformation parameter $%
\theta $ during the emission. This is a clear manifestation of the
Zeeman effect.\newline

Before closing this section, we present the Zeeman splitting for the
general case of $sl_{n}$ KdV- Burger's mapping as given by
\emph{fig.3}
%This is Figure 3
\begin{figure}[tbp]
\includegraphics[width=0.86\linewidth]{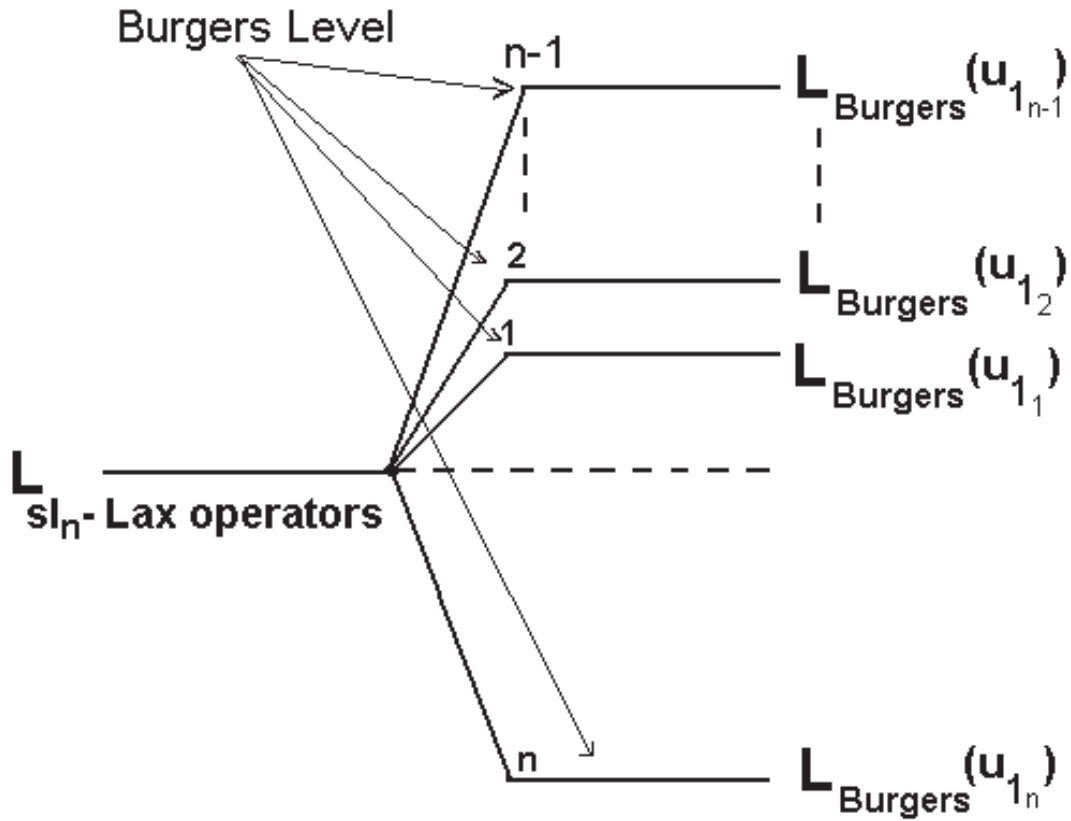}
\caption{\textbf{The $sl_n$-KdV-Burger mapping}} \label{fig3}
\end{figure}

\section{The BS construction: A review}

It's point out that gauge theories on noncommutative spaces
\cite{CDS, 5} are relevant to the quantization of D-branes in
background $B_{\mu \nu}$ fields \cite{6}. Such theories are shown
to be similar to ordinary gauge theory except that the usual
product of fields is replaced by a ``star
product'' that is well defined through the deformation parameter $%
\theta^{\mu \nu}$ which is an antisymmetric constant tensor. The
authors of \cite{BSus} show on a successful way, by means of
simple examples and of a convenient mathematical formulation, that
the noncommutativity is once again appearing as a rephrasing of
well known physics, namely the physics of the magnetic field.
\newline

The principal focus of this section is to recall a part of the BS
construction before considering an adaptation of their results to
our study. For this propose, we will begin by recalling this
construction based first on a simple quantum mechanical system, a
fundamental steps, before considering the string theory in the
presence of a D3-brane and a constant large $B_{\mu \nu}$ field.

\subsection{The classical model}

It deals with a system of two unit charges of opposite sign
connected by a spring with elastic constant $K$. The system is
embedded in a magnetic field $B$ in the regime where the Coulomb
and the radiation terms are negligible. The charges are associated
to particles of masse $m$ and charge $|q|=e$ and they are
localized at the coordinates $\vec{x_1}$ and $\vec{x_2}$ or in
component form $x_1^i$ and $x_2^i$. The Lagrangian of this system
is given by
\begin{equation}
{\cal L}={\cal L}_1+{\cal L}_2+{\cal L}_3
\end{equation}
The first term
\begin{equation}
{\cal L}_1 = \frac{m}{2} \left((\dot{x}_1)^2+(\dot{x}_2)^2 \right)
\end{equation}
is the kinetic energy of the charges.\newline The second term
\begin{equation}
{\cal L}_2 = \frac{B}{2} \epsilon_{ij} \left( \dot{x}_1^i x_1^j - \dot{x}%
_2^i x_2^j \right)
\end{equation}
describes the interaction of the charges with the magnetic field
$B$ and the last term
\begin{equation}
{\cal L}_3=-\frac{K}{2} (x_1-x_2)^2
\end{equation}
is known to be the harmonic potential between the charges.
\newline
\newline
{\bf Particular limit:}\newline {\em Concerning the kinetic term
of ${\cal L}$, it's assumed \cite{BSus} that the magnetic field
$B$ is so large that the available energy is insufficient to
excite higher Landau levels \cite{girvin}. So, the principal focus
will be on the lowest Landau level.}\newline

By virtue of this particular limit, the previous Lagrangian
reduces to
\begin{eqnarray}
\displaystyle{\ {\cal L} = \frac{B}{2} \epsilon_{ij} \left(
\dot{x}_1^i x_1^j - \dot{x}_2^i x_2^j \right) \: -\: \frac{K}{2}
(x_1-x_2)^2 }
\end{eqnarray}
From the classical mechanic's point of view, the canonical momenta
are given by
\begin{eqnarray}
\displaystyle{\
\begin{array}{l}
p_i^1 = \displaystyle{\ \frac{\partial {\cal L}}{\partial
\dot{x}_1^i} } = B
\epsilon_{ij} x_1^j, \\
p_i^2 = - B \epsilon_{ij} x_2^j;%
\end{array}
}
\end{eqnarray}
and the corresponding hamiltonian is deduced from the reduced
Lagrangian as follows
\begin{equation}
{\cal H} = 2K (\Delta)^2 = 2 K \left( \frac{P}{2B} \right)^2 =
\frac{K}{2B^2} P^2
\end{equation}
once the center of mass and relative coordinates $X$, $\Delta$ are
defined as:
\begin{eqnarray}
\displaystyle{\
\begin{array}{l}
\vec{X} = (\vec{x_1} + \vec{x_2}) /2 \\
\vec{\Delta} = (\vec{x_1} - \vec{x_2}) /2%
\end{array}
}
\end{eqnarray}
One learns from eq(29) that ${\cal H}$ is the hamiltonian of a
nonrelativistic particle with mass
\begin{eqnarray}
\displaystyle{\ M= \frac{B^2}{K} }
\end{eqnarray}
and center of mass momentum, conjugate to $X$, given by
\begin{equation}
P_i= 2 B \epsilon_{ij} \Delta^j =\frac{\partial {\cal L}}{\partial
\dot{X}^i}
\end{equation}
with
\begin{eqnarray}
\displaystyle{\ {\cal L} = 2 B \epsilon_{ij} \dot{X}^i \Delta^j -
2 K (\Delta)^2 }
\end{eqnarray}
As it's claimed by the authors of \cite{BSus}, this Lagrangian
exhibits a particular interest since an identical expression can
be extracted from string theory's point of view. The induced
message is that a quantum of NCYM can be described as an unexcited
string relating the particles of opposite charges inside a strong
magnetic field.\newline On the other hand, the interesting
consequence of the non locality as expressed by the behavior of
open string endpoints on D-brane world volume
\begin{equation}
\left[x^i, x^{j}\right]=i\theta^{ij}
\end{equation}
is the existence of dipole excitation \cite{BSus, 6} whose content
is expressed by $p_i=2B\epsilon_{ij}\Delta_{j}$ eq(32). One learns
from this relation that the spatial extension $|\Delta|$ of the
system is proportional to its momentum $|p|$ in the orthogonal
direction. This proportionality shows that the size of the system is
growing linearly with the momentum which is also equivalent to boost
the system along a direction while it will spread in the orthogonal
one leading simply to conclude that the particle is not any more
point-like.

\subsection{String theory's framework}

The principal task of this section is to show that string theory
analysis reproduces the same behavior of the Lagrangian
corresponding to the classical model. In fact, let's consider a
bosonic string theory in the
presence of a D3-brane oriented along the coordinates $x^{\mu },\mu =0,1,2,3$%
. We have also a background antisymmetric tensor field $B_{\mu \nu
}$ in the 1, 2 direction in light cone frame defined as follows
\begin{equation}
{\ x^{\pm }=x^{0}\pm x^{3}}
\end{equation}
with the usual light cone choice of world sheet time
\begin{equation}
\tau =x^{+}
\end{equation}

Focusing on the large limit for the magnetic field, namely
$B\longrightarrow \infty $, the string action can be reduced, upon
some rescalings and straightforward manipulations, to the
following form \cite{BSus}\newline

\begin{eqnarray}
\displaystyle{\ {\cal L}= \left[ -\frac{2 \Delta^2}{L} + \dot{y}
\epsilon \Delta \right] }
\end{eqnarray}
This Lagrangian shows a striking resemblance with the one of
classical mechanics, see eq(33).

\section{The BS construction adapted to $sl_n$ KdV-Burger's mapping}

We guess that a strong link should exists between the $sl_n$ KdV-
Burger's mapping with the interpretation provided through the
Zeeman effect, on one hand, and the BS construction from the other
hand.\newline

Indeed, the consideration of the authors of \cite{BSus} of the
simple classical model dealing with a pair of particles of
opposite charges embedded in a strong magnetic field and moving on
the infinite noncommutative $2d$ plane is of great interest
because of the remarkable similarity with the behavior of the
bosonic string theory in the presence of a $D3$-brane.

\subsection{The $sl_2$ KdV-Burger's mapping and the BS mechanical model}

We focus in what follows to use and adapt some important results
presented in the BS construction that we reviewed previously. Our
principal focus is to be able to interpret the Zeeman approach of
the $sl_n$ KdV Burger's mapping.\newline

For this reason, we consider the $2d$ phase-space ${\cal
F}_{\theta ,2}$ of
arbitrary functions $f(x,p)$ and $g(x,p)$ subject to the following Moyal $%
\theta $-NC rule
\begin{equation}
f\star g=\sum_{i,s=0}{\frac{\theta
^{s}}{s!}}(-)^{i}c_{s}^{i}(\partial _{x}^{i}\partial
_{p}^{s-i}f)(\partial _{x}^{s-i}\partial _{p}^{i}g),
\end{equation}%
such that $\{f,g\}_{\theta }=\frac{f\star g-g\star f}{2\theta }$
with $x$ and $p$ describing the space and momentum coordinates
respectively. \newline Next, we introduce a new space ${\cal
C}_{B,2}$ generated by the coordinates
$\vec{x_{1}}$ and $\vec{x_{2}}$ or in component form $x_{1}^{i}$ and $%
x_{2}^{i}$ \footnote{%
As it will be clear later, the lower indices in ${\cal F}_{\theta ,2}$ and $%
{\cal C}_{B,2}$ indicate the $\theta $ and $B$ predominance as
well as the order of the hierarchy and the number of emitted
Burger's levels respectively. Later on we will consider ${\cal
C}_{B,2}$ as being the BS space of opposite charges.}. These
coordinates are assumed to describe a pair of unit charges
$(q_{1},q_{2})$ of opposite sign $(q_{2}=-q_{1})$ in a magnetic
field $B$ in the regime where the Coulomb and the radiation terms
are negligible \cite{BSus}.\newline

We will give a series of propositions and ansatz with a principal
aim to install the link in question. The first step is given by
the following result:\newline
\newline
{\bf Proposition 3: NC $sl_{2}$-KdV's Phase ($B\rightarrow
0$):}\newline

Let's consider the {\em phase I} dominated by the noncommutative
$\theta $ deformation and characterized by the space ${\cal
F}_{\theta, 2}$ corresponding to the NC $sl_2$ KdV hierarchy with
the momentum NC Lax operator $p^2+u_{2}(x,t)$. From the point of
view of Zeeman effect this
hierarchy deals with a phase of degenerated single level, this is the NC $%
sl_{2}$-KdV Phase.\newline

{\bf Proposition 4: NC $sl_2$-Burger's Phase ($B\rightarrow \infty$):}%
\newline

In parallel, we consider a {\em phase II} with a magnetic field's
$B$ predominance and associated to the BS charge space ${\cal
C}_{B, 2}$ that we call, according to Zeeman effect, the
$sl_2$-Burger's phase. This phase is
characterized by a splitting of the degenerated NC $sl_2$ KdV single level $%
p^2+u_{2}(x,t)=(p+u_{1})\star (p-u_{1})$ to a set of $2=1+1$
Burger's levels under the effect of the strong magnetic field
$B$\newline
\newline
{\bf Ansatz}:\newline

By virtue of {\em Proposition 4}, the two emitted Burger's levels,
belonging to the space ${\cal C}_{B,2}$ in a strong magnetic
field's regime, are
associated to the opposite spin-one currents $u_{1_{_{1}}}=u_{1}$ and $%
u_{1_{_{2}}}=-u_{1}$ which are, in turn, in one-to-one
correspondence with the opposite BS charges $q_{1_{_{1}}}=q_{1}$
and $q_{1_{_{2}}}=-q_{1}$ respectively. We have
\begin{equation}
\begin{array}{lcl}
u_{1_{_{1}}}\equiv u_{1} & \Leftrightarrow & q_{1} \\
u_{1_{_{2}}}\equiv -u_{1} & \Leftrightarrow & -q_{1}%
\end{array}%
\end{equation}%
\newline
{\small {\it This ansatz gives a possibility to look at the non
degenerate Burger's levels, {\`{a}} la Zeeman, and the associated
conformal currents of opposite signs as being the two opposite
charges suggested by Bigatti and Susskind in their model
\cite{BSus}.\newline
\newline
Furthermore, the force of this ansatz is traced to the fact that
these two
charges as well as the two emitted Burger's levels exist in a phase ${\cal C}%
_{B,2}$ marked by a strong intensity of the magnetic
field.}}\newline
\newline
{{\bf Ansatz}: The (${\cal F}_{\theta ,2}$) - (${\cal C}_{B,2}$)
duality}
\newline
\newline
The space ${\cal F}_{\theta ,2}$ associated to the NC $sl_{2}$ KdV
single
level in a phase where $\theta $ predominate is dual to the space ${\cal C}%
_{B,2}$ of opposite BS charges in the strong magnetic field's
regime. The duality relation can be expressed formally as follows:
\begin{equation}
\begin{array}{lcl}
{\cal F}_{\theta ,2} & {\longleftrightarrow } & {\cal C}_{B,2} \\
\theta & {\longleftrightarrow } & {\theta ^{-1}}\sim B%
\end{array}%
\end{equation}%
\newline

\subsection{The general $sl_n$ KdV-Burger's mapping and the BS string-branes
models}

The previous results can be easily extended to higher order of the
NC KdV hierarchy. Before going into the general case, let's
consider the first non
trivial extension namely the third order class of the KdV hierarchy, we have%
\newline
\newline
{\bf Proposition 5: NC $sl_{3}$ Boussinesq's Phase ($B\rightarrow 0$):}%
\newline
Let's consider a {\em phase I} dominated by the noncommutative
$\theta $ deformation and characterized by the space ${\cal
F}_{\theta ,3}$ corresponding to the NC $sl_{3}$ Boussinesq
hierarchy with the momentum NC Lax operator $p^{3}+u_{2}\star
p+u_{3}$. From the point of view of Zeeman effect this hierarchy
deals with a phase of degenerated single level, this is the NC
$sl_{3}$-Bousinesq Phase. The degeneracy of the $sl_{3}$
Boussinesq single level in this phase is of order 3.\newline
\newline
{\bf Proposition 6: NC $sl_{3}$-Burger's Phase ($B\rightarrow \infty $):}%
\newline
\newline
In parallel, we consider a {\em phase II} with a magnetic field's
$B$ predominance and associated to the \textquotedblleft
charge\textquotedblright\ space ${\cal C}_{B,3}$. This phase is
characterized by a splitting of the degenerated NC $sl_{3}$ KdV
single level $p^{3}+u_{2}\star p+u_{3}=(p+u_{1})\star
(p+v_{1})\star (p-u_{1}-v_{1})$ to a set of $3=2+1$ Burger's
levels under the effect of the strong magnetic field $B$\newline
\newline
{\bf Ansatz}:\newline By virtue of {\em Proposition 6}, the three
emitted Burger's levels, belonging to the space ${\cal C}_{B,3}$
in a strong magnetic field's regime, are associated to the
following set of spin one currents:
\begin{equation}
\begin{array}{lcl}
u_{1_{_{1}}}=u_{1} &  &  \\
u_{1_{_{2}}}=v_{1} &  &  \\
u_{1_{_{3}}}=-u_{1}-v_{1} &  &
\end{array}%
\end{equation}%
\newline
\newline
In analogy with previous analysis, these three Burgers's currents $%
u_{1_{_{i}}},i=1,2,3$ are assumed to be associated to a set of
three charges
extending the BS opposite charges $(q_{1},-q_{1})$ in the following way:%
\newline
\begin{equation}
\begin{array}{lcl}
u_{1_{_{1}}}\equiv u_{1} & \Leftrightarrow  & q_{1} \\
u_{1_{_{2}}}\equiv v_{1} & \Leftrightarrow  & q_{2} \\
u_{1_{_{3}}}\equiv -u_{1}-v_{1} & \Leftrightarrow  & q_{3}=-q_{1}-q_{2} \\
&  &
\end{array}%
\end{equation}%
\newline
Note that the extended BS charges $q_{1},q_{2}$ are positives
while $q_{3}$
is negative\footnote{%
One may think of the positions of these three charges as forming a triangle.}%
.\newline
\newline
{{\bf Ansatz: The (${\cal F}_{\theta ,3}$) - (${\cal C}_{B,3}$) duality}}%
\newline
\newline
The space ${\cal F}_{\theta ,3}$ associated to the NC $sl_{3}$
Boussinesq
single level in a phase where $\theta $ predominate is dual to the space $%
{\cal C}_{B,3}$ of extended BS charges $q_{i},i=1,2,3$ in the
strong magnetic field's regime. We have: ${\cal F}_{\theta
,3}\longleftrightarrow {\cal C}_{B,3}$\newline
\newline
{{\bf Proposition 7: The $sl_{n}$ General case}}\newline
$\bullet $ The $n$ emitted Burger's levels, belonging to the space ${\cal C}%
_{B,n}$ in a strong magnetic field's regime, are associated to the
following
set of spin one currents: $%
u_{1_{_{1}}},u_{1_{_{2}}},...,u_{1_{_{n-1}}},u_{1_{n}}\equiv
-u_{1_{1}}-u_{1_{2}}-...-u_{1_{n-1}}$. \newline $\bullet $ The
$n=(n-1)+1$ Burgers's currents $u_{1_{_{i}}},i=1,2,...,n$ are
assumed to be associated to a set of $n$ charges extending the BS
opposite charges $(q_{1},-q_{1})$ in the following way:\newline
\begin{equation}
\begin{array}{lcl}
u_{1_{_{1}}} & \Leftrightarrow  & q_{1} \\
u_{1_{_{2}}} & \Leftrightarrow  & q_{2} \\
... & ... &  \\
u_{1_{_{n-1}}} & \Leftrightarrow  & q_{n-1} \\
u_{1_{_{n}}}=-\Sigma _{i=1}^{n-1}u_{1_{_{i}}} & \Leftrightarrow  &
q_{n}=-\Sigma _{i=1}^{n-1}q_{i} \\
&  &
\end{array}%
\end{equation}%
Note that the extended BS charges $q_{1},...,q_{n-1}$ belonging to
the space
${\cal C}_{B,n}$ are positives while $q_{n}$ is negative \footnote{%
By virtue of the Bigatti-Susskind construction, one may think of
the positions of these $n$ extended charges as forming an
hyperplan, and since the opposite charges of BS are shown to be
stretched by a spring ($\equiv $ open string), the charges of
${\cal C}_{B,3}$ in the triangle are assumed to
be stretched by a membrane or a 2-string while the $n$ charges of ${\cal C}%
_{B,n}$ are stretched by a $(n-1)$-brane. Note also the important
fact that all these systems of $k$ charges belonging to ${\cal
C}_{B,k}$, for arbitrary $k=1,2,...,n$ are neutral
systems.\newline }\newline
$\bullet $ One have the following $(\theta )-(\theta ^{-1}\sim B)$ duality: $%
{\cal F}_{\theta ,n}\longleftrightarrow {\cal C}_{B,n}$.\newline
\newline
{\bf Recapitulating diagram:}\newline
\begin{equation}
\begin{array}{lcccc}
&  &  &  &  \\
&  & mapping &  &  \\
& \Sigma _{2}^{(0,2)}/\Sigma _{2}^{(1,1)} & \hookrightarrow  &
(\Sigma
_{1}^{(0,1)},\Sigma _{1}^{(0,1)}) &  \\
& \Updownarrow  &  & \Updownarrow  &  \\
& {\cal F}_{\theta ,2} & \leftrightarrow  & {\cal C}_{B,2} &  \\
&  & duality &  &  \\
&  &  &  &
\end{array}%
\end{equation}%
\newline
We have introduced previously some consistent conventional
notations in the
Moyal momentum algebra's context \cite{BSed, BSed2} to describe the NC $%
sl_{2}$ KdV Lax operator ${\cal L}_{2}=p^{2}+u_{2}\equiv
(p+u_{1})\star (p-u_{1})$ through the following coset space
$\Sigma _{2}^{(0,2)}/\Sigma _{2}^{(1,1)}$. We proposed a mapping
between the $sl_{2}$ KdV system and the NC Burger's Lax
operators.\newline From the Zeeman effect's point of view, this
mapping is similar to a duality between the spaces ${\cal
F}_{\theta ,2}$ corresponding to the single degenerated KdV level
belonging to $\Sigma _{2}^{(0,2)}/\Sigma _{2}^{(1,1)}$ and ${\cal
C}_{B,2}$ describing the pair of emitted Burger's levels. We then
conclude that the space ${\cal C}_{B,2}$ of Burger's levels is the
same as the space of BS opposite charges (BS dipole).\newline
This analyse can be more generally extended to the order $n$ as follows%
\newline
\begin{equation}
\begin{array}{lcccc}
&  &  &  &  \\
&  & mapping &  &  \\
& \Sigma _{n}^{(0,n)}/\Sigma _{n}^{(n-1,n-1)} & \hookrightarrow  &
(\Sigma
_{1}^{(0,1)},...,\Sigma _{1}^{(0,1)}) &  \\
& \Updownarrow  &  & \Updownarrow  &  \\
& {\cal F}_{\theta ,n} & \leftrightarrow  & {\cal C}_{B,n} &  \\
&  & duality &  &  \\
&  &  &  &
\end{array}%
\end{equation}%

\section{Concluding Remarks:}

This work aims principally to present a new aspect of
noncommutative integrable models. We show that the spectrum
defined by the Lax pair with spectral parameter for the non
commutative deformation of certain $sl_{n}$ KdV integrable
hierarchies, namely the KdV and Burger's systems, exhibits a
degeneracy splitting reminiscent of the Zeeman effect.

The originality of this result and its natural aspect come from
the crucial use of the physical property $\theta \sim B^{-1}$
giving the known analogy between the non commutative parameter
$\theta $ and the magnetic field $B$.

The Zeeman effect in the present context seems to be a natural
incorporation for the following reasons:

\textbullet \qquad The $\theta $-Miura transformation is very
significant since its equivalent to a splitting of every $sl_{n}$
KdV hierarchy into$\ n$ different Burger's hierarchies.

The associated equation is given by: $L_{\{KdV\}}(u_{2})=p%
%TCIMACRO{\U{b2}}%
%BeginExpansion
{{}^2}%
%EndExpansion
+u_{2}=(p+u_{1})\star (p-u_{1}).$

\textbullet \qquad The KdV hierarchy described by the NC\ Lax operator $%
L_{\{KdV\}}(u_{2})$ and the current $u_{2}$ of conformal spin $2$ such that $%
u_{2}=-u_{1}^{2}-2\theta u_{1}^{\prime }$shows an explicit
dependence of the KdV current $u_{2}=u_{2}(\theta )$ in terms of
the non commutative $\theta $ parameter. Unlike the KdV current,
the spin $1$ Burger's current\ $u_{1}$
does not have this property since the associated NC\ Lax operator $%
L_{\{Burgers\}}(u_{1})$ does not admit a Lax operator, whose
conformal weight is integer, as a root. This is also due to the
fact that the quantities of non integer conformal weights are not
authorized in this framework.

\textbullet \qquad The NC Burger's current $u_{1}$ is regarded as
being a
fundamental current in term of which all the other currents of the $%
sl_{\{n\}}-KdV$ hierarchy are expressed.

\textbullet \qquad The Miura transformation, applied to $sl_{\{n\}}$%
noncommutative hierarchies, share with the famous Zeeman effect
the property of the splitting giving rise in turn to a couple of
phase of opposite magnetic field's dominance.

\textbullet \qquad In fact, from the Zeeman effect point of view, the NC $%
sl_{\{n\}}$ KdV hierarchy is associated to a phase $\theta\sim
B^{-1}$ where noncommutativity dominate, while the Miura
transformation leads to a splitting of the Burger's levels gives
rise to a new phase $B\sim \theta ^{-1} $ where the magnetic field
dominate.

To make concrete our idea, we have proceed by presenting a
graphical scenario similar to the Zeeman effect representation.
The essential of our representation for the general case, given in
Fig. 3, can be summarized as follows:

{\bf 1.} The original NC $sl_{n}$ Lax operator $L_{sl_{n}-KdV}$ is
represented by an horizontal line indexed by a multiplet of currents $%
(u_{2},u_{3},...u_{n})$. These currents indicate that the degree
of degeneracy inside this level is of order $n$. Note that the
degree of degeneracy is synonym of the highest degree in the non
commutativity whose parameter is $\theta$.

{\bf 2}. The splitting of $L_{sl_{n}-KdV}$, with respect to the
Miura
transformation, into a multiplet $%
(L_{Burg_{1}},,L_{Burg_{2}},...,L_{Burg_{n}})$ of NC Burger's
operators is
schematized by $n$ parallel lines leaving, from a vertex, the initial $%
sl_{n} $ KdV line.\newline

{\bf 3.} The position of the $n$ emitted Burger's lines depends on
the sign of spin $1$ Burger's currents. Uppers lines are for
example chosen to be associated to $L_{Burg}(\alpha u_{i}),\alpha
>0$ while lower lines are
associated to the negative sign of the NC Burger's currents, ie $%
L_{Burg}(\alpha u_{i}),\alpha <0$

{\bf 4.}. When the degeneracy of order $n$ is broken, the $n$ NC
Burger's levels are emitted. This emission procedure is given by
\begin{equation}
L_{Bouss}(u_{2},...,u_{n})\hookrightarrow
(L_{Burg}(u_{1}),...,L_{Burg}(u_{n}))
\end{equation}%
or equivalently
\begin{equation}
(p+u_{1})\star ...\star (p+u_{n})\hookrightarrow
((p+u_{1}),...,(p+u_{n}))
\end{equation}
{\textbf{Acknowledgments}}\\
{\small E.H.S. and M.B.S. would like to thank the Abdus Salam
International Center for Theoretical Physics (ICTP), where this work
was done during our visit on July-August 2006. M.B.S. acknowledge
the high energy section and its head S. Randjbar-Daemi for the
invitation. The authors acknowledge valuable contributions of
OEA-ICTP in the context of NET-62 and present also special
acknowledgements to the Associateship Scheme of
ICTP}.\\


\begin{thebibliography}{99}
\bibitem{snyder} {\footnotesize H.S. Snyder, Quantized space-time, Phys.
Rev. 71 (1947) 38; The electromagnetic field in quantized
space-time, Phys. Rev. 72 (1947) 68. }

\bibitem{witten} {\footnotesize E. Witten, Noncommutative geometry and
string field theory, Nucl. Phys. B 268 (1986) 253. }

\bibitem{1} {\footnotesize A. Connes, Noncommutative geometry, Academic
Press 1994. %\cite{Connes:1997cr}
}

\bibitem{CDS} {\footnotesize \ A.~Connes, M.~R.~Douglas and A.~S.~Schwarz,
%``Noncommutative geometry and matrix theory: Compactification on tori,''
JHEP {\bf 9802} (1998) 003 [arXiv:hep-th/9711162].
%%CITATION = HEP-TH 9711162;%%
%\cite{Douglas:1997fm}
}

\bibitem{5} {\footnotesize \ M.~R.~Douglas and C.~M.~Hull,
%``D-branes and the noncommutative torus,''
JHEP {\bf 9802} (1998) 008 [arXiv:hep-th/9711165].
%%CITATION = HEP-TH 9711165;%%
%\cite{Sheikh-Jabbari:1998ac}
}

\bibitem{6} {\footnotesize \ M.~M.~Sheikh-Jabbari, ``Super Yang-Mills theory
on noncommutative torus from open strings %interactions,''
Phys.\ Lett.\ B {\bf 450} (1999) 119 [arXiv:hep-th/9810179].
%%CITATION = HEP-TH 9810179;%%
%\cite{Seiberg:1999vs}
}

\bibitem{7} {\footnotesize \ N.~Seiberg and E.~Witten,
%``String theory and noncommutative geometry,''
JHEP {\bf 9909} (1999) 032 [arXiv:hep-th/9908142].
%%CITATION = HEP-TH 9908142;%%
}

\bibitem{books} {\footnotesize J. Madore, ``An Introduction to
Noncommutative Geometry and its Physical Applications,'\ Cambridge
University Press, 1999.\newline G. Landi, ``Introduction to
Noncommutative Spaces and their Geometries'', Springer-Verlag,
97.\newline M.B. Green , J.H. Schwarz and E. Witten, ``Superstring
theory,'\ Cambridge Univ. Press, 1987. \newline J. Polchinski,
``String Theory,'\ Cambridge Univ. Press, 1998.\newline C.V.
Johnson, ``D-Branes,'\ Cambridge Univ. Press, 2003.
%\cite{Boulahoual:2002yh}
}

\bibitem{BSed} {\footnotesize \ A.~Boulahoual and M.~B.~Sedra, ``The Moyal
momentum algebra applied to theta-deformed 2d conformal models and
KdV-hierarchies,'' Chin.\ J.\ Phys.\ {\bf 43}, 408 (2005)
[arXiv:hep-th/0208200]. %%CITATION = HEP-TH 0208200;%%
}

{\footnotesize %\cite{Boulahoual:2005zu}
}

\bibitem{BSed2} {\footnotesize \ A.~Boulahoual and M.~B.~Sedra, ``The Moyal
momentum algebra,'' A.\ J.\ Math.\ Phys.\ {\bf 2}(2005)111.
%%CITATION = 00451,2,111;%%
%\cite{Dafounansou:2005hd}
}

\bibitem{DES} {\footnotesize \ O.~Dafounansou, A.~El Boukili and
M.~B.~Sedra, ``Some aspects of noncommutative integrable systems
grave a la
Moyal,'' arXiv:hep-th/0508173. %%CITATION = HEP-TH 0508173;%%
%\cite{Sedra:2005fg}
}

\bibitem{Sed} {\footnotesize \ M.~B.~Sedra, ``Moyal noncommutative
integrability and the Burger-KdV mapping,'' Nucl.\ Phys.\ B {\bf
740} (2006)
243 [arXiv:hep-th/0508236]. %%CITATION = HEP-TH 0508236;%%
%\cite{Bigatti:1999iz}
}

\bibitem{BSus} {\footnotesize \ D.~Bigatti and L.~Susskind,
%``Magnetic fields, branes and noncommutative geometry,''
Phys.\ Rev.\ D {\bf 62}, 066004 (2000) [arXiv:hep-th/9908056].
%%CITATION = HEP-TH 9908056;%%
}

\bibitem{girvin} {\footnotesize S.~M.~Girvin and Terrence Jach, Formalism
for the Quantum Hall effect: Hilbert space of analytic functions,
Phys.~Rev.~B 29, 5617 (1984) }

\bibitem{hep-th/0110191} {\footnotesize %\cite{Ho:2001aa}
P.~M.~Ho and H.~C.~Kao, \textquotedblleft Noncommutative quantum
mechanics
from noncommutative quantum field %theory,''
Phys.\ Rev.\ Lett.\ {\bf 88} (2002) 151602 [arXiv:hep-th/0110191].
%%CITATION = HEP-TH 0110191;%%
}
\end{thebibliography}
\end{document}